\documentclass[%
  aip,
 amsmath,assume,
 multicol,multirow,
 reprint,%
 ]{revtex4-1}
\usepackage{multirow}
\usepackage{graphicx}
\usepackage{amsfonts}
\usepackage{graphicx}
\usepackage{dcolumn}
\usepackage{bm}
\usepackage[mathlines]{lineno}
\usepackage{cases}
\usepackage[utf8]{inputenc}
\usepackage[T1]{fontenc}
\usepackage{mathptmx}
\usepackage{etoolbox}

\makeatletter
\def\@email#1#2{%
 \endgroup
 \patchcmd{\titleblock@produce}
  {\frontmatter@RRAPformat}
  {\frontmatter@RRAPformat{\produce@RRAP{*#1\href{mailto:#2}{#2}}}\frontmatter@RRAPformat}
  {}{}
}%

\draft 

\begin{document}


\title{Modeling Turbulent and Self-Gravitating Fluids with Fourier Neural Operators} 



\author{Keith Poletti}
\email{kepo6935@utexas.edu}
\altaffiliation{Corresponding Author}
\affiliation{Oden Institute, University of Texas at Austin, U.S.A}
\author{Stella S.~R. Offner}
\affiliation{Department of Astronomy, University of Texas at Austin, U.S.A}
\author{Rachel A. Ward}
\affiliation{Department of Mathematics, University of Texas at Austin, U.S.A}


\date{\today}

\begin{abstract}

Neural Operators (NOs) are a leading method for surrogate modeling of partial differential equations. Unlike traditional neural networks, which approximate individual functions, NOs learn the mappings between function spaces. While NOs have been predominantly tested on simplified 1D and 2D problems, such as those explored in prior works, these studies fail to address the complexities of more realistic, high-dimensional, and high-dynamic range systems. Moreover, many real-world applications involve incomplete or noisy data, which has not been adequately explored in current NO literature. In this work, we present a novel application of NOs to astrophysical data, which involve high-dynamic range projections into an observational space. We train Fourier NO (FNO) models to predict the evolution of incomplete observational proxies with density variations spanning four orders of magnitude. We demonstrate that FNOs can predict the effects of unobserved dynamical variables. Our work lays the groundwork for future studies that forecast direct astronomical observables. 

\end{abstract}

\pacs{}

\maketitle 

\section{Introduction}

NOs constitute a key pillar in the rapidly advancing field of Scientific Machine Learning. NOs were first described in 1995 by Chen\cite{Chen1995}, where the authors proved the universal approximation theorem for nonlinear operators with neural networks; however, they have only become computationally feasible in the past few years \cite{DeepONet,kovachki2021neural}. NOs differ from standard neural networks in that they learn a nonlinear operator between infinite-dimensional function spaces. They are invariant to different grid resolutions. Because they approximate the underlying operator, NOs have become a leading method for surrogate modeling of Partial Differential Equation (PDE) solutions. This strength makes NOs a potentially significant tool for solving inverse problems that require a large number of PDE solves\cite{LiFNO, olearyroseberry2023derivativeinformed, takamoto2023pdebench}. Recent NO architectures can be divided into several main network types that have spawned multiple variants: DeepONets\cite{DeepONet}; kernel integral neural operators, e.g., graph neural operators \cite{kovachki2021neural}; FNOs\cite{LiFNO}; and PCA-Nets\cite{PCANet}. In this paper, we focus on FNO-based methods because prior studies \cite{ dehoop2022costaccuracy, gupta2022towards, li2022learning, takamoto2023pdebench} highlight their accuracy in predicting the properties of turbulent flows as well as their ability to capture the energy power spectrum across many orders of magnitude.

To date, FNOs have been primarily applied to terrestrial problems and have been directly compared to numerical solvers at the same resolution. Common examples include the vorticity formulation of the incompressible Navier-Stokes equations, Darcy flow, and shallow water equations \cite{gupta2022towards, LiFNO, takamoto2023pdebench}. Other methods have been developed to predict turbulent flow with high Reynolds numbers \cite{li2022learning}, assuming a global attractor exists. Benchmark studies have compared the accuracy of FNOs against other neural surrogates on sub-sonic ($\mathcal{M} =0.1$) and sonic ($\mathcal{M}=1.0$) flows with two or three dimensions and multiple scalar and vector inputs \cite{takamoto2023pdebench,gupta2022towards}. While FNOs performed well, these applications are generally proof-of-concept problems involving simplifications of complex natural dynamics. Recently, NOs have also been applied to magnetohydrodynamics (MHD) problems \cite{FNOMHD,multiFNOMHD,PINOMHD}, successfully forecasting plasma observations in a Tokamak fusion reactor. Alternate FNO variants have been proposed to learn multiphase flows \cite{wen2022ufno}, dynamics on a spherical surface \cite{bonev2023spherical}, and interactions between input channels via Clifford Algebras \cite{brandstetter2023clifford}. One notable high-dimensional real-world use case is in weather forecasting applications, in which various FNO architectures\cite{FourCastNet2022,bonev2023spherical} have efficiently predicted long-range atmospheric dynamics. However, FNOs have yet to be trained for flow conditions characteristic of astrophysical systems, which are described by highly compressible, high-mach number flows that span a large dynamic range in density and spatial scale. Alternative approaches, such as Physics Informed Neural Networks, incorporated the dynamical equations and improved the performance for highly turbulent systems\cite{zhao2025lesnetslargeeddysimulationnets}. These physics-informed methods have been trained on coarse-grid data and were fine-tuned on a small amount of fine-grid data\cite{wang2024closuremodelslearningchaoticsystems}. Generative modeling has also shown success in predicting invariants of turbulent flows\cite{molinaro2025generativeaifastaccurate}. 

Magnetized plasma dynamics underlie a broad range of phenomena, ranging from combustion to nuclear fusion control. Within the field of astronomy, MHD turbulence represents the fundamental dynamics of interstellar gas, stellar winds, and galactic dynamics \cite{McKee_2007, Hopkins_MHD_2015, beattie2024magnetizedcompressibleturbulencefluctuation, Grudic_2021,hopkins2024forgediiformationmagneticallydominated, Weber_Davis_1967}. 
Interstellar gas, particularly the cold gas associated with star-forming regions, is weakly ionized, often strongly magnetized, and highly compressible, distinguishing it from the plasmas characteristic of terrestrial applications, which are generally strongly ionized and incompressible. Numerical MHD simulations, which are carried out with a variety of public codes, have been essential in building our current theory of astrophysical systems \cite{Athena++2020, Fryxell_2000, Hopkins_2015, Hopkins_MHD_2015}. However, numerical models contain significantly more detailed information than data taken by direct telescope observations, which are constrained by resolution, radiative transfer effects, and characteristic evolutionary time scales that exceed the human lifespan. 

Telescope observations effectively impose a noisy, nonlinear transformation on the underlying physical conditions. Observations project physical quantities, e.g., magnetic field strength and gas density, onto a 2D ``plane of sky" image. The 3D gas velocity is sampled through atomic and molecular spectra, which only reflect velocity motions on the axis toward or away from the observer \cite{Beaumont_2013}.  Since these data are incomplete by nature and it is impossible to verify observations experimentally, numerical simulations enable hypothesis testing and constraints on relevant physics \cite{Haworth_2018}. Observations evolve on timescales much longer than the human lifespan, and thus, these data are essentially static.
Due to the incomplete physical information underlying the observations, MHD simulations, which require detailed, well-specified initial conditions, cannot be used to predict evolution from an observed ``snapshot". However, recent advances in machine learning\cite{FNOMHD, multiFNOMHD} have enabled predictions in the \textit{observational space}, wherein simulations mapped to the space of directly measurable parameters act as training sets\cite{Xu_2020a, Xu_2020b,xu2023predicting, Xu_2023mag, Xu_2023}. These NO's were directly trained on images from inside the MAST Tokamak fusion reactor, which provides experimental evidence for learning a non-unique operator. Additionally, FourCastNet \cite{FourCastNet2022} and the Spherical-FNO\cite{bonev2023spherical} train FNO variants to forecast weather variables with a reduced set of parameters from the ERA5 dataset\cite{ERA5}. The dynamics of these experiments evolve much faster than those of astrophysical systems, allowing the construction of time-series datasets. Thus, for this proof-of-concept, we derive our datasets from numerical simulations instead of relying on experiments.


In this paper, we examine the performance of FNO architectures for predicting dynamical evolution in three astrophysically relevant scenarios. Our first application models a spherical gas cloud collapsing under its own gravity. This setup is spherically symmetric, so it is effectively ``one-dimensional," since only the radius evolves; however, we train the models on data defined in Cartesian coordinates, which do not assume spherical symmetry. This setup tests NO performance in a new context that spans a dynamic range of three orders of magnitude. Accurately predicting the evolution of systems characterized by large dynamic ranges is crucial for applying ML techniques to astrophysical problems. In the second application, we simulate three-dimensional, supersonic hydrodynamic turbulence and then train NOs to forecast the observational proxies. In the final application, we model supersonic magnetized turbulent flow, including the magnetic field as an additional latent variable, to evaluate how well NOs perform when trained on data without explicit knowledge of the magnetic field, i.e., when the number of unknown and unconstrained parameters increases significantly. Each application increases the amount of missing information in the training data to test NO performance in domains with reduced physical constraints, e.g., as in telescopic observations.

\section{Data Generation} In this study, we train three FNO architectures to learn a nonlinear operator mapping from an observation space to itself. Although producing training data that is a projection into the low-dimensional observational space is straightforward, no well-defined inverse mapping exists from this space to the original high-dimensional function space. 

\subsection{Training Data} 

Our data are derived by performing an ``observational" mapping on the simulation outputs. We tailor the standard operator learning formulation with the additional observational constraint as the following problem setup. Let $\mathcal{U}(\Omega)$ be some Banach space where $\Omega \subset \mathbb{R}^d$. A nonlinear operator $A: \mathcal{U}\to \mathcal{U}$ is the forward solution of some PDE, i.e., if $f_t \in \mathcal{U} $, then $f_{t+1}:= A(f_t) $. Let $\mathcal{V}(K)$ be the observation space corresponding to $\mathcal{U}$ defined on $K \subset \Omega$. Let $\phi:\mathcal{U} \to \mathcal{V}$ be a non-invertible mapping. We assume that there exists a nonlinear operator $G:\mathcal{V}\to \mathcal{V}$  that forward propagates the observations, i.e. $\phi(f_{t+1}) := G(\phi(f_{t})).$ We approximate with the NO, $\Tilde{G}_{\theta} \approx G$. The NO weights $\theta \in \Theta$ are optimized with respect to the following minimization problem:
\begin{equation}
    \min_{\theta \in \Theta} \ \mathbb{E}_{t} \left[ \frac{\| \Tilde{G}_{\theta} (\phi(f_t)) - \phi(f_{t+1}) \|_\mathcal{V}}{ \| \phi(f_{t+1})\|_{\mathcal{V}}}\right]. 
    \label{min_prob}
\end{equation}
\begin{figure}
    \centering
    \includegraphics[width=0.50\textwidth]{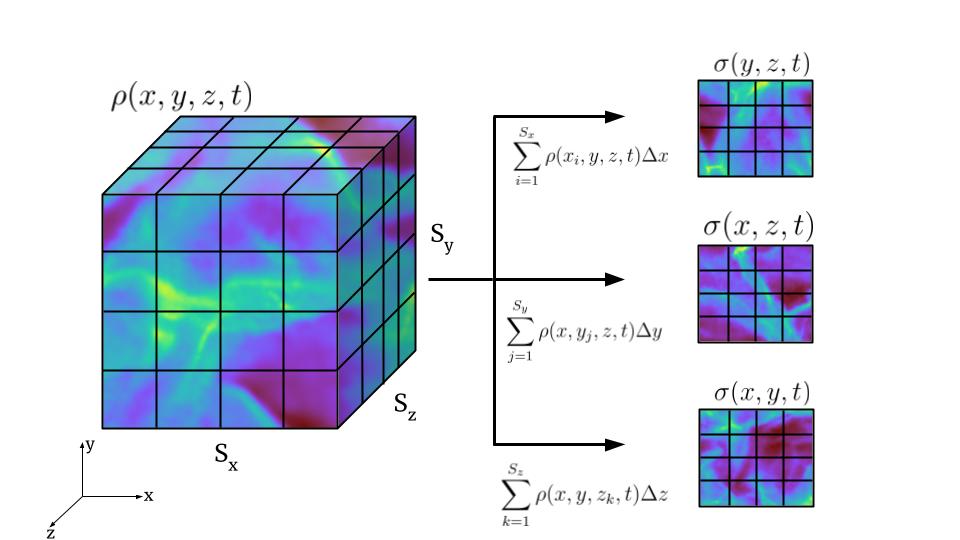}
    \caption{A schematic illustrating the approach used to generate proxy density observations. The 3D simulation gas density is converted from unstructured cells into a 3D uniform grid with resolution $S_x \times S_y \times S_z$ where $S_x=S_y=S_z=64$ in our case. The discretized proxy-observation mapping, $\mathbb{R}^{S_x} \times \mathbb{R}^{S_y} \times \mathbb{R}^{S_z} \ni \rho \mapsto \sigma \in\mathbb{R}^{S_y} \times \mathbb{R}^{S_z}$, projects the density along an axis by summing over that axis. The proxy-observation mapping for every snapshot is repeated for the \textbf{x, y,} and \textbf{z} axes.}
    \label{fig:dataProjection}
\end{figure}

In this work, we use hydrodynamic numerical simulations to create a training set that serves as a proxy for actual telescope observations. These proxies depend on the ``line-of-sight" vector, which points from the observer to the object. The plane, which is normal to the ``line-of-sight" vector, is called the ``plane-of-sky" and represents the image seen by an observer. Our simulated data are produced by 
solving the coupled PDEs describing gas dynamics
as described in the following sections. The resulting simulated flows are projected by summing along the ``line-of-sight" axis as a simplified proxy for observations, as shown in Figure \ref{fig:dataProjection}. We use the \textit{yt} package \cite{Turk_2010} to project the density, $\rho$, and the velocity, $\mathbf{u}$,
where each velocity component is defined as $ \mathbf{u}:=\begin{bmatrix}u&v&w \end{bmatrix}^T$, in the two orthogonal directions to the ``line of sight." These projections produce three functions: an integrated density, $\sigma(x,y,t)$, and a 2D plane of the sky vector for the velocity $(\Bar{u},\Bar{v})$, which are the input and output channels for our models. Each projection sums along one of the three axes as in equation \eqref{eq:proj}  for the $\hat{\mathbf{z}}$ axis:
\begin{subequations}
    \label{eq:proj}

    \begin{equation}
        \sigma(x,y,t) =  \sum_{k=1}^{S}\rho(x,y,z_k,t) \Delta z \label{subeq:proj_rho}
    \end{equation}
    \begin{equation}
        \Bar{u}(x,y,t) =  \sum_{k=1}^{S}  u(x,y,z_k,t), \label{subeq:proj_u}
    \end{equation}
    \begin{equation}
        \Bar{v}(x,y,t) =\sum_{k=1}^{S}   v(x,y,z_k,t) \label{subeq:proj_v},
    \end{equation}
\end{subequations}

where $S$ is the number of spatial grid points, and $\Delta z$ is the spatial grid resolution. For every snapshot (except in the spherical collapse case), we repeat this projection for the $\mathbf{\hat{x}}$ direction, $\mathbf{\hat{y}}$ direction, and $\mathbf{\hat{z}}$ direction to increase the number of training samples. Since the spherical collapse problem is symmetric, we only project along the $\mathbf{\hat{z}}$ plane. 

Spectroscopic observations of collapsing gas, as found in the dense cores of star-forming regions, detect emission that probes the entire gas column, thus comprising the ``column density" and the gas motion along the line-of-sight; therefore, we adopt the projected gas density, Eq. \eqref{subeq:proj_rho}, and velocity in the plane-of-sight, Eqs. \eqref{subeq:proj_u} and Eqs. \eqref{subeq:proj_v},  as our training quantities. We format the training data on a 64 $\times$ 64 grid for all simulations.

\subsection{Spherical Collapse} 

The most straightforward system we consider is a self-gravitating sphere of initially uniform density gas. Gravitationally driven collapse is a common occurrence in astrophysical systems and is a necessary precursor to phenomena like star and planet formation.  Our setup is spherically symmetric, such that the gas contracts radially. This first proof-of-concept experiment tests FNO accuracy when the input data are a 2D projection of a 3D dynamical system and when the data span an extensive dynamic range. The equations of motion are the gravitational potential, Eq. \eqref{grav}, the continuity equation, Eq. \eqref{cont}, and the inviscid isothermal Euler equation, Eq. \eqref{euler},:
\begin{align}
    &\nabla^2 \phi(x,y,z,t) = 4 \pi \rho(x,y,z,t),  \label{grav} \\
    &\frac{\partial \rho}{\partial t} + \nabla \cdot (\rho \mathbf{u}) = 0, \label{cont}\\
    &\rho \left(\frac{\partial \mathbf{u}}{\partial t} +(  \mathbf{u} \cdot \nabla)  \mathbf{u}\right) = - \nabla p \label{euler},
\end{align}
where $\phi$ is the gravitational potential, $\rho$ is the mass density, and $\mathbf{u}$ is the velocity vector. While Gizmo solves these equations in a Lagrangian frame, we present them in the Eulerian frame, because we train the models in this frame. The simulations use a meshless finite-mass solver, a unique Lagrangian Godunov-type method that solves the equations with a moving cell distribution \cite{Hopkins_2015}. We initialize the gas at rest within a domain of $[-l, l]^3 $pc$^3$  (1 pc = $3.086 \times10^{16}$ m) and adopt periodic boundary conditions where $l=0.125$ pc. We assume the gas is isothermal with a fixed temperature of $T=20 \ K$ and begins with no initial velocity. The total system mass controls the initial density, and we run simulations with 30 initial masses between $M_0 \in [0.48 M_{\odot}, 10 M_{\odot}]$ while keeping the initial radius constant.

We divide the simulation snapshots into sets of 20 steps with a constant $\Delta t=0.18$ kilo-years, where we use the first four timesteps as inputs to the NO and the last four as a test set to evaluate the predictive accuracy. The training dataset contains 450 time series at various stages of the spherical collapse. 

Here, we learn the operator that maps observational proxies for the density and velocity in the $(-l,l)^2$ ``plane of the sky" in the time interval $ [ 0, 4 \Delta t]$ to their future state at $[15 \Delta t, 19 \Delta t]$,
\begin{equation}
\Tilde{G}_\theta : C\left([0, 4 \Delta t]; H^1((-l, l)^2;\mathbb{R})\right) \to C \left([15 \Delta t, 19\Delta t]; H^1((-l, l)^2;\mathbb{R})\right), 
\end{equation}
where $l$ is the length of the grid and $H^1$ is the standard Sobolev space with bounds on the first gradient,  i.e.,  
\[H^1((-l, l)^2;\mathbb{R})  \colon =   \left\{ f \in  L^2((-l, l)^2;\mathbb{R}) : \ \| f\|_{L^2} + \| \nabla f \|_{L^2} < \infty \right\}. \]
In terms of each of our three channels,  we learn a channel-wise mapping \[ (\sigma,\Bar{u},\Bar{v})\rvert_{(-l, l)^2 \times [0,4 \Delta t]} \mapsto (\sigma,\Bar{u},\Bar{v})\rvert_{(-l, l)^2 \times [14 \Delta t, 19 \Delta t]}.\]

\subsection{Turbulent Flow} 

Next, we examine the case of a 3D highly compressible turbulent flow without gravity or magnetic fields. This turbulent dataset explores the application of NOs to data with structures spanning a large dynamic range, characteristic of the plasma properties in astronomical systems. These data are far more disordered than those of the spherical collapse case. 

We generate the training data by solving the 3D, inviscid flow, isothermal Euler equations in  Eqs. \eqref{cont} and \eqref{euler}.
The simulation domain is a 3D cube with sides of length 2.5 pc ($l=1.25$ pc) and periodic boundary conditions. Each simulation follows $3.3 \times 10^4$ fixed-mass cells with a total mass of $ 600 M_{\odot}$. To create turbulent conditions, we mix the gas for $3 t_{\rm cross}$, where the crossing time $t_{\rm cross}$ is defined as the length of the domain divided by the average velocity dispersion, $v_{rms}$. This step ensures that the turbulence reaches a quasi-steady state \cite{Lane_2021}. 
Unlike most previous works that study incompressible or weakly compressible plasmas \cite{FNOMHD, PINOMHD}, we model a highly compressible, supersonic gas with  $\mathcal{M} = v_{rms}/c_s = 12$.  

To produce the training set, we project the resulting simulation data along three axes to create observational proxies. 
To increase diversity in our dataset, we include simulations with 36 different initial random seeds, where we fix the gas mass and velocity dispersion to reduce the data variance. The final training set comprises 9,600 time series with 20 timesteps per series with a timestep of $\Delta t = 8.29$ kilo-years. We split the time series similarly to the spherical collapse case: for each set of 20 consecutive time steps the first four are given as an input to the network, and the last four are used to evaluate the output network prediction. 
As in the spherical collapse model, we learn the forward mapping from $ (\sigma, \Bar{u}, \Bar{v})\rvert_{(-l,l)^2 \times [0,4 \Delta t]} \mapsto (\sigma, \Bar{u}, \Bar{v})\rvert_{(-l,l)^2 \times [15 \Delta t, 19 \Delta t]}.$


\subsection{Turbulent Magnetohydrodynamics} Our next test case examines the FNO performance in predicting the evolution of supersonic MHD turbulence. These equations, Eqs. \eqref{cont}, \eqref{con_mhd}, and \eqref{con_mag}, expand on the previous Euler equations, Eqs. \eqref{cont} and \eqref{euler}, by incorporating magnetic fields: 
\begin{align}
        &\rho \left(\frac{\partial \mathbf{u}}{\partial t} +( \mathbf{u} \cdot \nabla) \mathbf{u}\right) = - \nabla p - \frac{1}{4 \pi}\mathbf{B} \times (\nabla \times \mathbf{B}) \label{con_mhd}\\
        &\frac{\partial }{\partial t}\left(\frac{\mathbf{B}}{\rho} \right) + \mathbf{u} \cdot \nabla \left(\frac{\mathbf{B}}{\rho} \right)  = \left( \frac{\mathbf{B}}{\rho} \cdot \nabla\right) \mathbf{u}, \label{con_mag}   
\end{align}
where  $\mathbf{B}$ is the magnetic field.
We solve the equations in the ideal MHD limit, where the gas and magnetic field are assumed to be perfectly coupled, i.e., no ambipolar diffusion or Ohmic dissipation occurs. 

We generate the simulation data similarly to the non-magnetized turbulence case, and likewise, the setup assumes the domain is a 3D cube with sides of length 2.5 pc ($l=1.25$ pc) and periodic boundary conditions.  The initial mass, virial parameter, and resolution are the same as in the turbulent fluid experiment, and training data is taken from times $ t > 3 t_{\rm cross}.$ The initial magnetic field is uniform across the three spatial dimensions and is given by $\mathbf{B}=\begin{bmatrix}10^{-4} & 10^{-4} & 10^{-4}\end{bmatrix} G$. 

To assess the sensitivity of the result to the underlying magnetic field - an astrophysically poorly constrained quantity - we only train the network on density and velocity, as described above. By excluding the magnetic field, we challenge NOs to learn and predict the dynamics of the observables when crucial physical quantities are not observed. As before, we learn the forward mapping from $ (\sigma, \Bar{u}, \Bar{v})\rvert_{(-l,l)^2 \times [0,4 \Delta t]} \mapsto (\sigma, \Bar{u}, \Bar{v})\rvert_{(-l,l)^2 \times [15 \Delta t, 19 \Delta t]}$ with $\Delta t =8.29$ kilo-years.

\section{Methods}
\subsection{Fourier Neural Operators} 

The FNO approximates a mapping between two function spaces and comprises three stages. First, the input is lifted into a higher dimensional space with a lifting layer, $ \mathcal{L}:f(t,x) \to g_0(t,x)$. Once in the higher dimensional space, the data pass through multiple Fourier Layers, $g_i \mapsto g_{i+1}$. Each Fourier layer convolves the input in the Fourier domain and applies a skip connection transformation in the Euclidean domain. The \texttt{neuralop} library provides multiple options for the skip connection, which we adopt as a hyperparameter. The skip connection options we test are the identity mapping, a linear pointwise transformation, and a channel-weighted soft-gating.
The Fourier Layer operation is given by:
\begin{equation}
    \label{FourierLayer}
    g_{i+1}(t,x) = \psi( A g_{i}(t,x) + \mathcal{F}^{-1}(R_\phi (\mathcal{F}g_i))(t,x)),
\end{equation}
where the Fourier convolution, $R_{\phi}$, is applied to the Fourier transformation of the previous layer's output, $\mathcal{F}g_i$. The result is then inverse Fourier transformed back to the Euclidean domain, and the Euclidean skip connection, $A$, is applied to the previous layer's output. The outputs are summed together, and the activation function, $\psi$, completes the Fourier Layer. The \texttt{neuralop} architecture optionally applies an additional two-layer multilayered perception after each Fourier Layer and adds the Euclidean skip connection \cite{kossaifi2023multigrid}. The last layer, $\mathcal{P}$, projects the Fourier layer output to a lower dimensional space. 

We employ two FNO variations to approximate the nonlinear operator: the standard FNO-3D architecture and an autoregressive FNO-3D architecture. Both methods were proposed for modeling data in two spatial dimensions and one temporal dimension\cite{LiFNO}. The standard FNO-3D architecture convolves across both space and time and treats time as a third dimension. This method trains faster since it involves only a single application of the network.
These models truncate the higher frequency modes, and the output loses small-scale structure with repeated network applications. 
In the FNO-3D training step, we input five evenly spaced timesteps and train the network to predict five timesteps 15 timesteps later.

The auto-regressive FNO-3D architecture convolves the input data both spatially and temporally as well and is recurrently applied to predict further in time. When training this model, the networks forecast for smaller timesteps and require more memory and longer training times. Recursive training requires multiple applications to predict further in time.
When training the auto-regressive FNO-3D model, we input only the first four timesteps, then predict the next four timesteps sequentially. These timesteps are recurrently fed through the network until the twentieth timestep is reached. 

\subsection{U-Shaped Neural Operator}

U-Shaped Neural Operators (UNOs) combine the dynamic layer sizes of U-Nets and the integral operators of FNOs\cite{rahman2023uno}. A UNO architecture begins with a lifting layer, $L$, and ends with a projection layer, $P$, similar to the standard FNO architecture. The intermediate layers comprise a sequence of nonlinear integral operators that encode into a contracted domain while increasing the dimension of the co-domain. Following the encoder, a separate decoding sequence of nonlinear integral operators maps into an expanded domain with a lower dimensional co-domain. This encoding and decoding process reduces the memory required per layer, enabling the construction of networks with more layers. 

In this work, we train a 3D UNO that convolves across two spatial dimensions and one time dimension. The UNO architecture requires more memory compared to the standard FNO-3D architecture, so we adopt a non-autoregressive training setup to conserve memory. 

\section{Results}
We report the Normalized Root Mean Square Error (N-RMSE) for the testing data, $\delta_f$ in Table\ref{RMSE_tab}. We calculate the test set mean of the N-RMSE for each field, the average of those field-wise errors, and the maximum error field-wise average of the test set.
The N-RMSE defined for an arbitrary proxy variable $f$, is given by: 
\begin{equation}
    \delta_{f} = \frac{1}{T} \sum_{t=1}^{T} \sqrt{ \frac{ \sum_{i=1}^S \sum_{j=1}^S \left(\Tilde{G}_{\theta}(f(x_i,y_j,t)) -f(x_i,y_j,t) \right)^2}{ \sum_{i=1}^S \sum_{j=1}^S \left(f(x_i,y_j,t) \right)^2}}, 
    \label{n-rmse}
\end{equation}
which averages the grid-wise squared error between predicted proxy, $\Tilde{G}_{\theta}(f)$, and the true proxy, $f$, in the $S \times S$ grid where the proxy, $f$, can be  $\sigma$, $\Bar{u}$, or $\Bar{v}$. We normalize the error with respect to the root mean squared spatial signal, because the data span a large dynamic range in values between fields. We average N-RMSE  across all discrete $T$ timesteps in the target operator, where $T=4$. 

In addition to the NO architectures, we compare the error of the Identity operator by taking the normalized root mean square error between the inputs and outputs. Since our data do not exhibit much evolution based on visual inspection, the identity operator sets a baseline for the effectiveness of the surrogate models.

\subsection{Spherical Collapse Model Predictions} 

\begin{figure*}
\begin{center}
  \includegraphics[width=\linewidth]{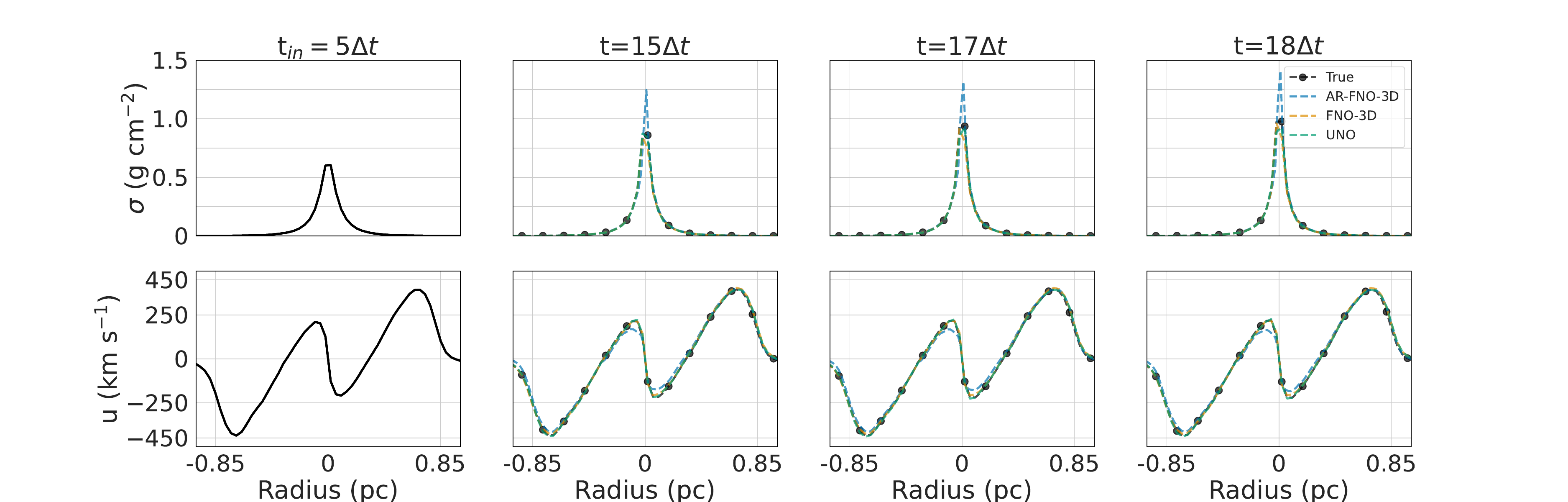}
  \caption{Cross section slices for a collapsing sphere, where time increases from left to right. The first column is the final timestep input into the model, while each sequential column depicts the prediction at the first ($t=15\Delta t$), third ($t=17\Delta t$), and final ($t= 18 \Delta t$) output times.
  This run has a sphere mass of $0.63 M_{\odot}$ and timestep of $\Delta t =$0.18 kilo-years. The rows from top to bottom show the slice of integrated density and the slice of projected velocity.}\label{Grav_density} 
  \end{center}
\end{figure*}

Figure \ref{Grav_density} shows that the AR-FNO-3D model, the FNO-3D model, and the UNO model accurately predict the short-term evolution of the collapsing sphere. 
We show a one-dimensional slice across the center to display the differences between each model. The evolution displayed in Figure \ref{Grav_density} occurs toward the end of the collapse, at which point most of the change occurs in the core center. 
The AR-FNO-3D model peaks at the highest density, while the FNO-3D and UNO 
match the true maximum density. The projected velocity reflects a performance similar to that of the integrated density. Both the FNO-3D and UNO 
exhibit similar performance while the AR-FNO-3D does not predict the velocity accurately in the center.
During the cloud collapse, the projected density spans four orders of magnitude, $[ 10^{-4}, 1] \frac{g}{cm^2}$. 

The N-RMSEs given in Table \ref{RMSE_tab} indicate that all three models outperform the Identity mapping for all reported values. The FNO-3D model performs the best for the field average N-RMSE. 
In all cases, the N-RMSE increases at the final timesteps in the simulations. These late stages change rapidly compared to the initial stages and thus are underrepresented in the training dataset. 

\subsection{Turbulent Flow Model Predictions} 
\begin{figure*}[!hbt]
    \centering
    \includegraphics[width=\linewidth]{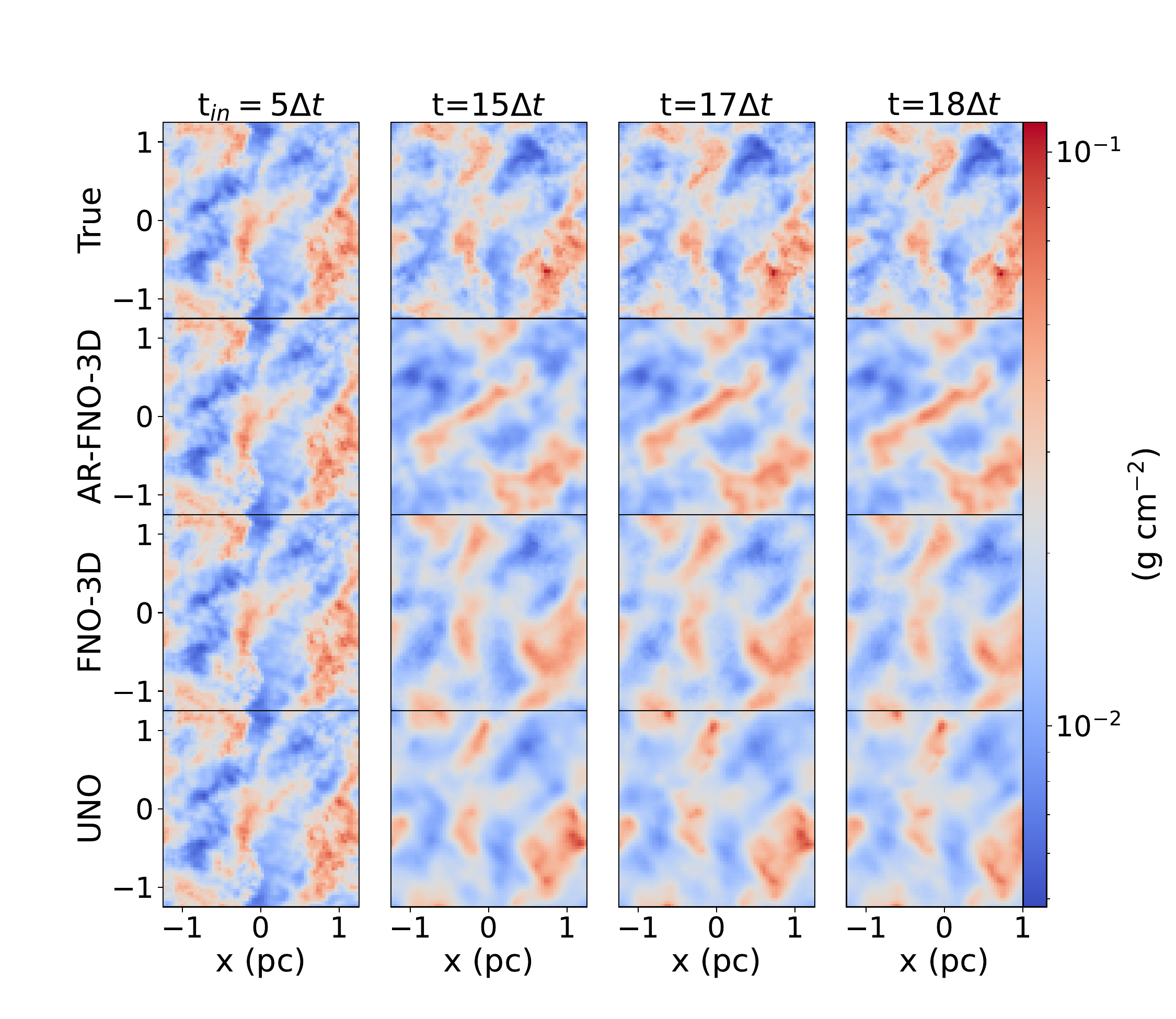}
   \caption{Temporal evolution
   of the integrated density after the system reaches a quasi-steady turbulence state. The timestep is $\Delta t=$ 8.29 kiloyears. The first column shows the final timestep input into the model, while each subsequent column depicts the first ($t=15\Delta t$), third ($t=17\Delta t$), and final ($t= 18 \Delta t$) output times. The rows from top to bottom are the Gizmo solution, and the auto-regressive FNO-3D,  FNO-3D, and UNO predictions, respectively.}
   \label{turb_dens}
\end{figure*}

The compressible flow produces features across all spatial scales. The top row of Figure \ref{turb_dens} 
shows that the ground truth integrated density contains both large structures approximately 1 pc in length and small filamentary structures on the order of 0.1 pc. The AR-FNO-3D prediction, shown in the second row of Figure \ref{turb_dens}, elongates the filaments and merges them into larger structures that do not exist in the true distribution. While the FNO-3D predictions preserve the filament structures, the filaments do not span the full range of densities. 
The UNO predictions, shown in the bottom row of Figure \ref{turb_dens},  visually match the true distribution the best. However, all three models smear the smaller scale features. 

\begin{table*}[htbp]
    \centering
    \caption{Comparison of the N-RMSE between each neural network method. The third column indicates the number of parameters in the network. The last five columns summarize the errors as calculated from Equation \ref{n-rmse}: N-RMSE for integrated density, $\delta_\sigma$, N-RMSE for the first velocity component, $\delta_u$, and second velocity component, $\delta_v$, the average of the errors $\delta_{avg}$, and the maximum average error for all test data, $\delta_{max}$. The bold numbers are the best-performing model for the metric and experiment.}
    \begin{ruledtabular}
    \begin{tabular}{llllllll}
        \hline
        \textbf{Simulation}&\textbf{Method}&\# Parameters& $\delta_{\sigma}$ & $\delta_u$ & $\delta_v$ & $\delta_{avg}$ & $\delta_{max}$  \\ \hline
        \multirow{4}*{\textbf{Spherical Collapse}}&Identity& N/A&  0.089&  0.218&  0.217&  0.174&  0.360 \\
        &AR-FNO-3D& 4.16e+08&  0.089&  0.126&  0.104&  0.106&  0.257 \\
        &FNO-3D& 1.78e+08&  \textbf{0.029}&  \textbf{0.062}&  0.068&  \textbf{0.053}&  0.111 \\
        &UNO& 1.43e+09&  0.054&  0.062& \textbf{0.062}&  0.059&  \textbf{0.095} \\ \hline
        \multirow{5}*{\textbf{Turbulence}}&Identity& N/A&  0.413&  0.404&  0.384&  0.400&  0.622 \\
        &AR-FNO-3D& 1.55e+08&  0.269&  0.250&  0.245&  0.255&  0.487 \\
        &FNO-3D& 1.83e+08&  \textbf{0.246}&  0.261&  0.246&  0.251&  \textbf{0.385} \\
        &UNO& 2.11e+07&  0.286&  \textbf{0.218}&  \textbf{0.231}&  \textbf{0.245}&  0.702 \\ \hline
        \multirow{4}*{\textbf{Magnetohydrodynamics}}
        &Identity& N/A&  0.435&  0.268&  0.310&  0.338&  0.591 \\
        &AR-FNO-3D& 1.55e+08&  0.306&  0.174&  0.209&  0.229&  0.362 \\
        &FNO-3D& 6.54e+07&  0.234&  0.132&  \textbf{0.156}&  0.174&  0.275 \\
        &UNO& 1.40e+07&  \textbf{0.231}&  \textbf{0.132}&  0.157&  \textbf{0.173}&  \textbf{0.276} \\
    \end{tabular}
    \end{ruledtabular}
    \label{RMSE_tab}
\end{table*}

\begin{figure*}[!hbt]
    \centering
    \includegraphics[width=0.9\linewidth]{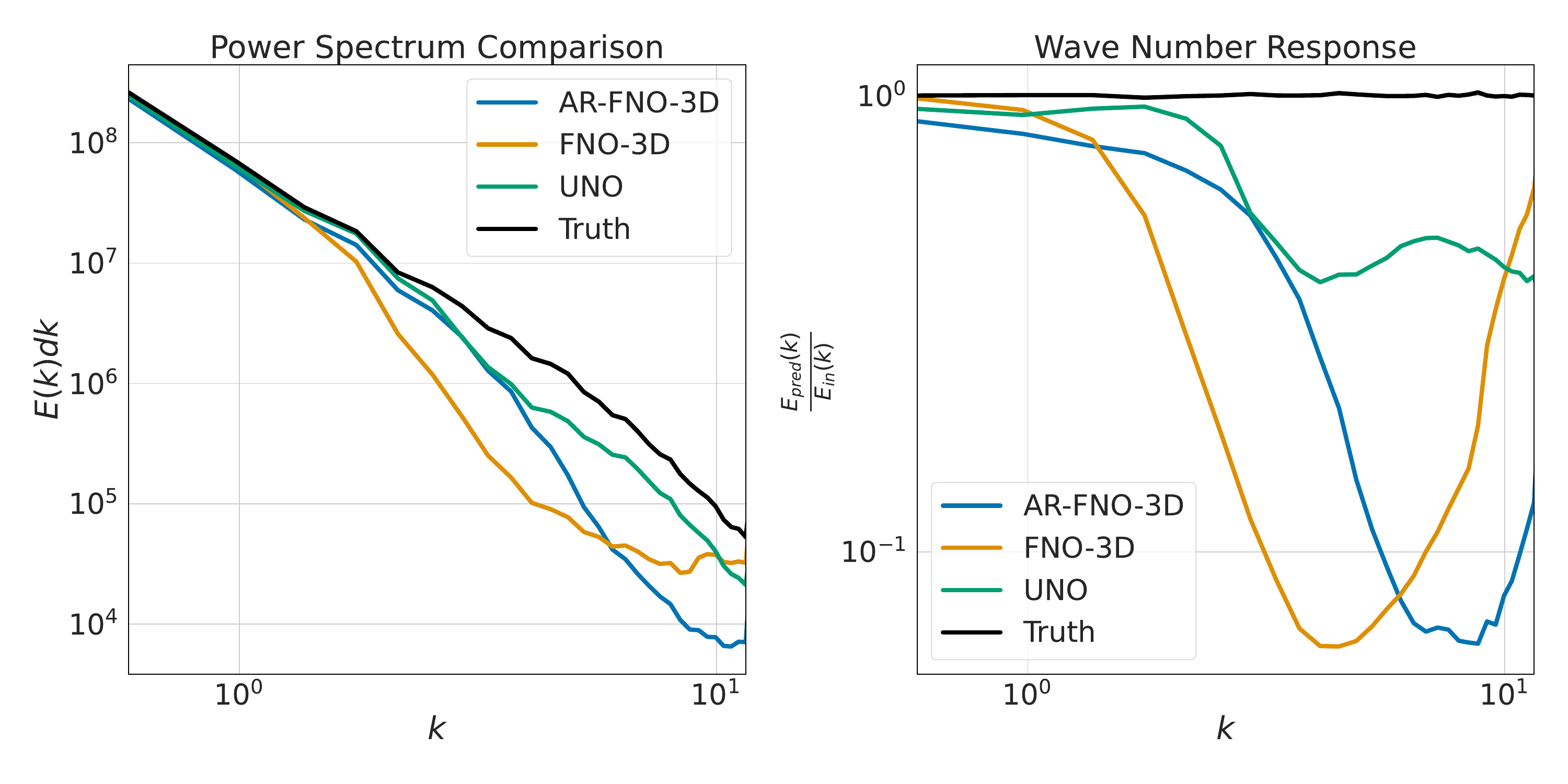}
    \caption{Left: Power spectrum of the hydrodynamic turbulence model predictions where $k$ is the corresponding wavelength. Right: Ratio of the power predicted to the input power where $k$ is the corresponding wavelength.}
    \label{fig:Turb-Spec}
\end{figure*}

Figure \ref{fig:Turb-Spec} shows the power spectrum for the hydrodynamic turbulent flow problem. All models perform best at low wave numbers, i.e., they all reproduce large structures in the flow. At high wave numbers,  the power spectrum of all three models decays faster than the true power spectrum, which corresponds to the filtering of small spatial structures. These spectra reflect the averaging effect and missing small-scale structure illustrated by Figure \ref{turb_dens}.
The UNO model preserves the most input power across all wave numbers.
For wave numbers above four, the slopes of the FNO predictions steepen significantly, indicating the significant loss of small-scale structure. 
Both FNO-3D models filter the higher wave number modes 
more than expected given the number of modes used in the spectral convolutions: 40 modes for the auto-regressive FNO-3D model and 46 modes for the FNO-3D model. The UNO model performs significantly better, despite only using 17 modes in the spectral convolution.
   
\subsection{Magnetized Turbulent Flow Model Predictions} 
\begin{figure*}[!hbt]
\centering
        \includegraphics[width=0.9\linewidth]{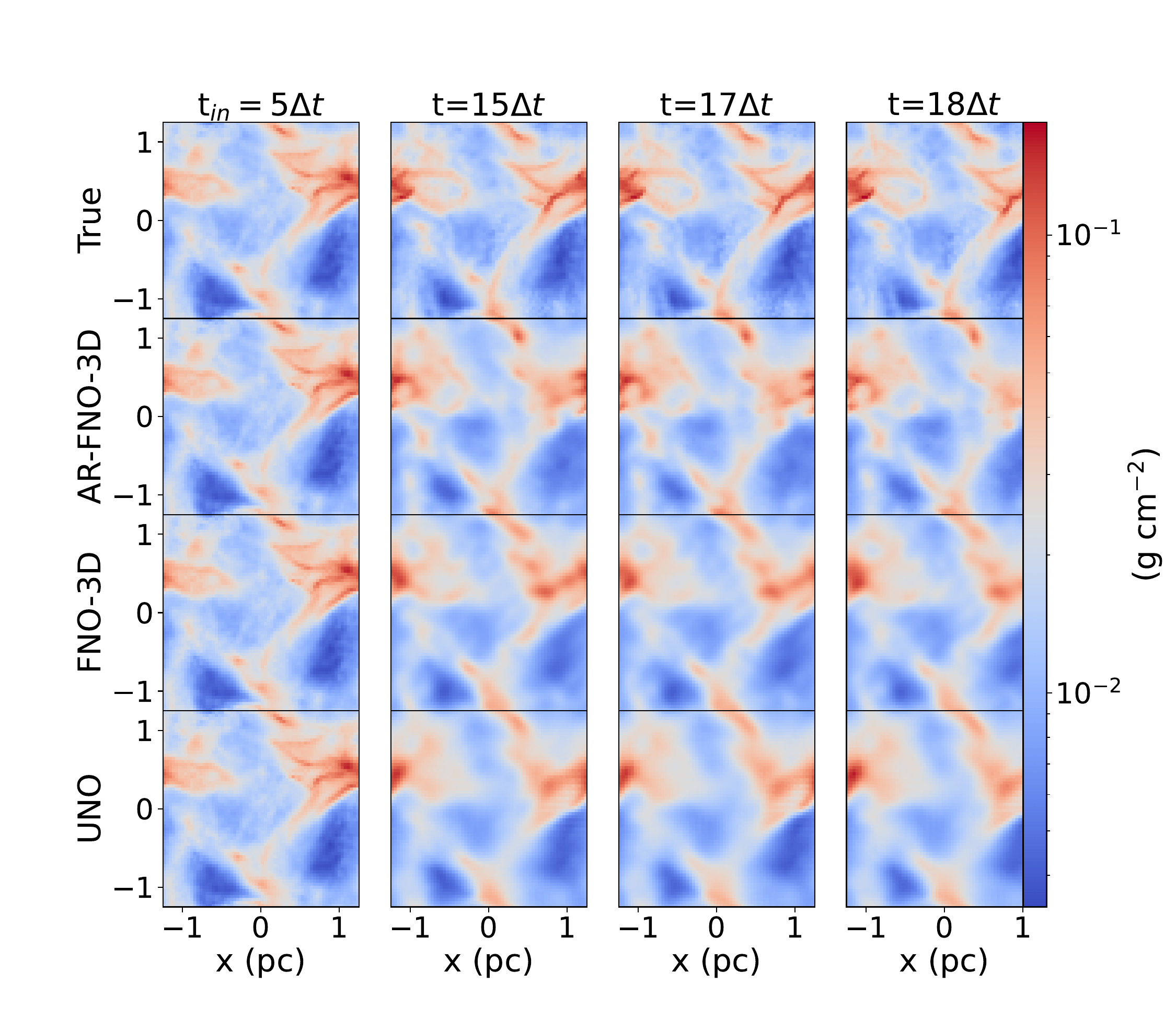}
   \caption{Temporal evolution
   of the integrated density after the magnetized fluid reaches a quasi-steady turbulence state. The timestep is $\Delta t=$ 8.29 kiloyears. The first column shows the final timestep input into the model, while each subsequent column depicts the first ($t=15\Delta t$), third ($t=17\Delta t$), and final ($t= 18 \Delta t$) output times. The rows from top to bottom are the Gizmo solution, the auto-regressive FNO-3D, FNO-3D, and UNO predictions, respectively.}
  \label{MHD_dens}
\end{figure*}
The magnetized turbulence is smoother than non-magnetized turbulence because the magnetic field adds additional pressure and reduces the density contrast across shocks, effectively reducing small-scale structure. Consequently, despite having more latent variables, the evolution of the magnetized turbulence is easier to capture. The bottom panels of Figure \ref{MHD_dens} show that the Fourier Layer-based models accurately predict the short-term time series of the projected data and recover most of the large-scale structure. However, as before, all models smear fine-scale structures. In Table \ref{RMSE_tab}, the UNO model exhibits the best performance, closely followed by the FNO-3D model. Compared to the setup without magnetic fields, the N-RMSE is lower, likely due to the smoother turbulent flow. The AR-FNO-3D generates the most detailed substructures in the high-density filaments, but those substructures do not match the substructures in the true distribution.

In comparison, the UNO and FNO-3D predictions average the substructure of the high-density filaments. Compared to the true density, this averaging blurs the entire domain and removes most small-scale structures. In addition, the small-scale filaments have shorter dynamical timescales and evolve more rapidly compared to the large filaments. The fine structure present in the AR-FNO-3D predictions is likely due to the auto-regressive predictions on shorter timescales than the UNO model and the FNO-3D model. Since the UNO model and FNO-3D models ignore the intermediate timesteps, the small-scale dynamics evolve too fast for the models to predict accurately, and these models instead average the larger structures.

\begin{figure*}[!hbt]
    \centering
    \includegraphics[width=0.9\linewidth]{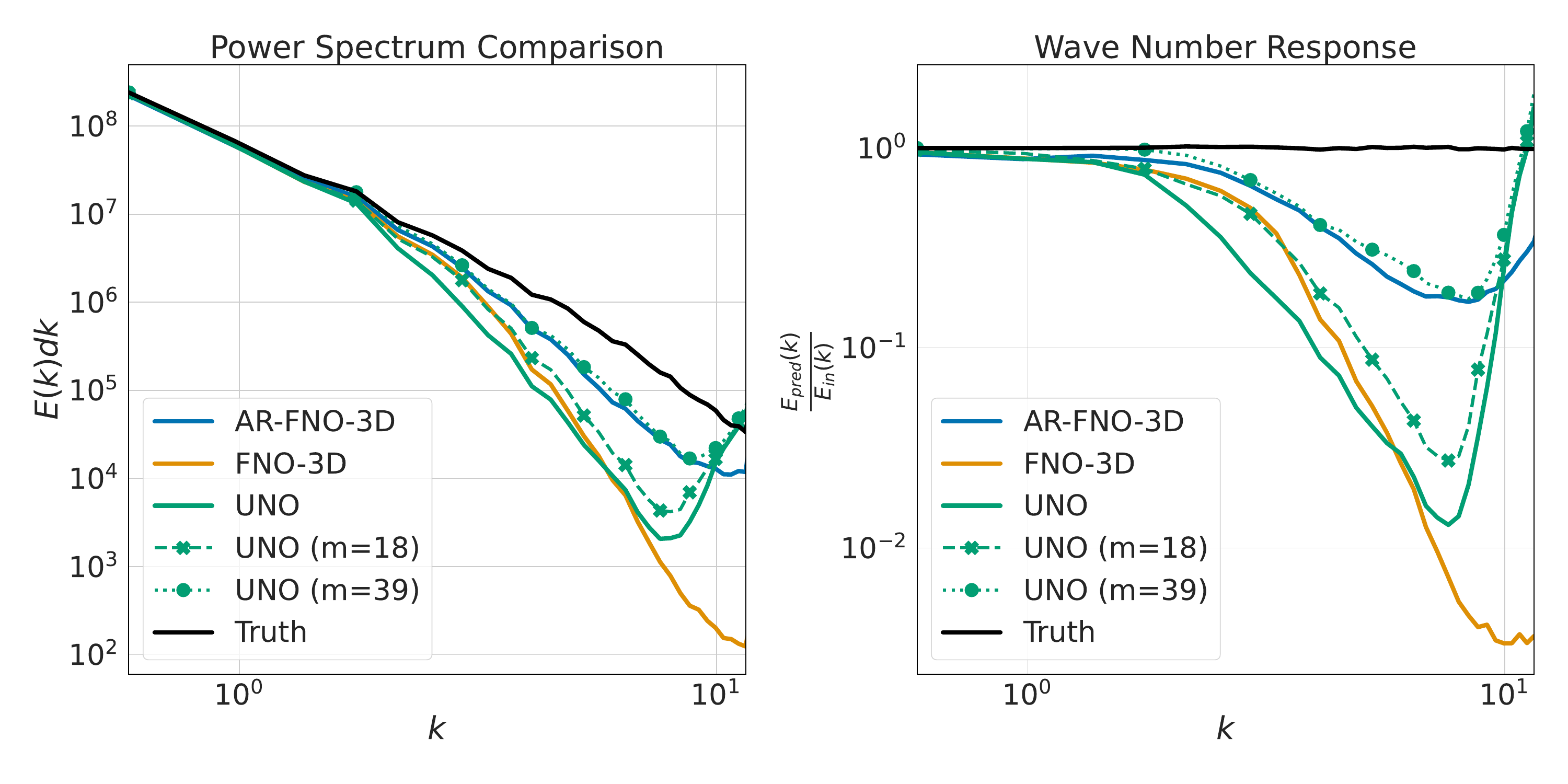}
    \caption{Left: Power spectrum of the MHD turbulent models. Right: Ratio of the power predicted to the input power. 
   }
    \label{fig:MHD-Spec}
\end{figure*}

Figure \ref{fig:MHD-Spec} depicts the MHD turbulent power spectrum and the wave number response for each model. Here, the AR-FNO-3D model most closely follows the slope of the true power spectrum. 
The FNO-3D and UNO models, however, miss most of the small-scale flow structure,  
and their power spectra quickly decline at wave numbers beyond four. The UNO model acts like a stop-gap filter by removing intermediate frequencies and possibly aliasing higher frequencies. Here, the AR-FNO-3D model retains 39 modes in the spectral convolution, while the FNO-3D and UNO models retain 18 and 12 modes, respectively. 

We test the impact of the number of modes by retraining the UNO model, which has the fastest decaying power spectrum, with the same number of modes as the AR-FNO-3D (m = 39) and FNO-3D (m = 18). In both cases, the retrained power spectrum closely follows the model with the equivalent number of modes. The retrained models increase around a wave number of nine. These mode differences mostly explain the different performances in the power spectra. We chose the best-performing models based on N-RMSE, but the lowest N-RMSE does not guarantee that a model's power spectrum will be more accurate.

We include figures for the velocity predictions for each problem in Appendices \ref{Spherical_App},  \ref{Turb_App}, and \ref{MHD_App}. 

\section{Discussion and Conclusions}

In this study, we demonstrate the ability of FNOs and UNOs to produce temporal predictions based on observational proxies. These proxies have limitations, such as missing key parameters, a low-resolution grid, and projection from three dimensions to two dimensions, that are characteristic of telescope observations. Our differs from the common approach of directly comparing a NO architecture to the numerical solver with the same grid resolution, same parameters, and same operators\cite{FNOMHD,multiFNOMHD, PINOMHD}. While the prior comparison is imperative for benchmarking novel architectures, additional tests are needed to benchmark performance on the application to observational data.
We examine the performance of these architectures for highly turbulent systems with Mach numbers an order of magnitude larger than previously reported. We present three cases with varying levels of missing information. In the case modeling spherically symmetric gravitational collapse, our results show that NOs, trained on projected data, accurately predict the evolution over a dynamic range spanning four orders of magnitude. No prior works to date have investigated fluid flows with a dynamic range larger than two orders of magnitude. We find that the FNO-3D and UNO models perform comparably in predicting the evolution of magnetized plasmas, even when no magnetic field information is included in the training. The autoregressive FNO-3D model preserves the spectral slope of the MHD turbulent flow. However, the overall error for each tested model decreased, likely due to the smoothing influence of the magnetic field on the flow.

Table \ref{RMSE_tab} shows that all the NO models achieve a N-RMSE on the order of $10^{-1}$. These N-RMSEs are comparable to the forward prediction accuracy for sonic ($\mathcal{M} =1.0$) inviscid flows reported in PDEBench\cite{takamoto2023pdebench}. For example, a three-dimensional computational fluid dynamics (3DCFD) simulation predicted with a FNO-3D model achieves N-RMSEs of $2.4 \times 10^{-1}$ for turbulent velocity initial conditions and $3.7 \times 10^{-1}$ for random velocity initial conditions (see Table 13 from PDEBench). These two PDEBench experiments test the prediction accuracy of the FNO-3D model on a full 3D grid with a size of 64$^3$. Unlike our training set, however, the PDEBench 3DCFD problem learns the operator for a simpler system of equations, in which the training data contain the full information required to solve the system using a classical PDE solver. 

We show that NOs do not require complete information covering all of the physical quantities that dictate the evolution of magnetized turbulent flow. However, the prediction accuracy is significantly degraded compared to models with complete information\cite{FourCastNet2022,bonev2023spherical}. 
This suggests that either larger models, more training data, or different architectures are needed. Training on more complete data would undoubtedly improve the accuracy of the predictions but is problematic from an astronomy perspective given the limits on those available data. Despite their limited accuracy, the tested models could reduce the search space for inverse problems due to their relative speed. The final inverse problem solution would still require high-fidelity PDE solvers in the reduced space. 


The present work lays the groundwork for future studies that include gravity, turbulence, and magnetic fields altogether, which are required to train models to accurately predict on actual telescope observations. Predicting on observations requires additional datasets, which must include radiative transfer and noise from various sources. Including these extra physical parameters reduces the distance between training data distribution and the observational data distribution. We identify several weaknesses in current NO architectures that should be addressed in future work, including memory limitations that constrain model size, and therefore, the prediction accuracy. Memory-efficient methods could allow for models with more Fourier modes or wider networks, resulting in lower error and better resolution of fine structure\cite{kossaifi2023multigrid,white2023speeding}. More sophisticated NO variants
suggest a promising approach to achieve better accuracy in inferring high-dynamic range input variables, a key requirement for effectively modeling astrophysical systems\cite{brandstetter2023clifford, multiFNOMHD}. NOs, like PDE solvers, perform best when tailored to specific problems. As such, the Local Neural Operator (LocalNO) is a promising new approach to solve the smoothing problem, which combines nonlocal Fourier layers with local differential and/or local integral layers\cite{liuschiaffini2024neuraloperatorslocalizedintegral}. To reduce the degeneracy in the operator, conservation laws may also be useful in constraining the space of possible solutions and improving the accuracy of the models. These constraints should be used only for fine-tuning since physics-informed losses are known for instability when training\cite{wang2024closuremodelslearningchaoticsystems}. 

\section*{Acknowledgments}
Keith Poletti and Rachel Ward acknowledge financial support in part by the Air Force Office of Scientific Research Multidisciplinary Research Program of the University Research Initiative FA9550-19-1-0005, National Science Foundation's Division of Mathematical Sciences-1952735, National Science Foundation's Institute for Foundations of Machine Learning grant 2019844, National Science Foundation's Division of Mathematical Sciences-N2109155, and National Science Foundation  2217033 and National Science Foundation 2217069. Stella Offner acknowledges support from the Oden Institute through a Moncrief Grand Challenge Award, support from National Science Foundation grant 2107942, and NASA grant 80NSSC23K0476. Keith Poletti thanks Qing Zhu for insightful conversations.
\section{Author Declarations.}
\subsection*{Conflict of Interest}
The authors have no conflicts to disclose.
\subsection*{Author Contributions} \textbf{Keith Poletti:} Conceptualization (supporting), Data Curation (lead), Formal Analysis (equal), Methodology (equal), Software (lead), Validation (lead), Visualization (equal), Original Draft Preparation (lead), Review and Editing (supporting).
\textbf{Stella Offner:} Conceptualization (lead), Methodology (equal), Project Administration (equal), Resources (equal), Supervision (equal), Validation (supporting), Visualization  (equal), Review and Editing (lead). \textbf{Rachel Ward:} Conceptualization (supporting), Formal Analysis (equal), Funding Acquisition (lead), Methodology (supporting), Project Administration (equal), Resources (equal), Supervision (equal), Review and Editing (supporting).
\subsection*{Data Availability}
The data that support the findings of this study are openly available in Zenodo at \href{https://doi.org/10.5281/zenodo.14796870}{https://doi.org/10.5281/zenodo.14796870}, reference number zenodo.14796870. The Code is available at \href{https://github.com/KPoletti/NeuralOperatorStarForm}{https://github.com/KPoletti/NeuralOperatorStarForm}

%
%

%


\appendix
\section{Training Procedure}
Our models were fine-tuned for each architecture via a random search on various hyperparameters, e.g., learning rate, number of Fourier modes, layer width, number of layers, type of skip-connection, and weight decay, for 100 epochs using \texttt{Wandb}. We varied the training loss function between the relative $L^2$ loss function and the relative $H^1$ loss function\cite{olearyroseberry2023derivativeinformed} although we calculated the gradients using the central difference method. We then trained the best-performing networks for 500 epochs. Table \ref{RMSE_tab} reports the number of parameters used in the best-performing models. We optimized our models with Adam \cite{kingma2017adam} with Cosine Annealing \cite{loshchilov2017sgdr} learning rate.

All models were trained on the Texas Advanced Computing Center's Lonestar6 supercomputer. We trained on a single A100 GPU for the hyperparameter tuning. We trained the optimal models with either PyTorch's Distributed-Data-Parallel on three A100 GPUs or on a single H100 GPU, depending on availability. This multi-GPU method spawns an identical model on each GPU and then synchronizes the networks at a checkpoint epoch, which was the twentieth epoch in this case. 

Before training, we normalized the data to a unit Gaussian distribution. We split the data between 80\% training, 10\% validation, and 10\% testing. Validation data were tested each epoch while the testing data were reserved until all training was completed.

\section{Cost-Accuracy Assessment.} 
Neural network scale-up is critical for performing more accurate predictions; however, scaling can lead to overfitting, require more computing power, and diminish the benefits neural networks offer compared to numerical solvers. To this end, we present a cost-accuracy assessment for the MHD model, which is representative of all three cases. 
To narrow the scope of comparison, we compare how the width of the hidden channels affects the N-RMSE. We 
compare network widths of 16, 32, 64, 96, or 128 (unless otherwise stated). 
We optimize the models using a smaller set of hyperparameters, namely the batch size and learning rate, 
by using a grid search. 
If the operator is sufficiently smooth, the N-RMSE should decrease as the width increases until the models start over-fitting the training data\cite{dehoop2022costaccuracy}. However, our operators are non-smooth due to the discontinuities caused by shocks and the projection into observational space. Similarly to the non-smooth problems in prior works, we expect to see limited improvements in performance 
with respect to the width of the network.

\begin{figure}[!hbt]
    \centering
    \includegraphics[width=\linewidth]{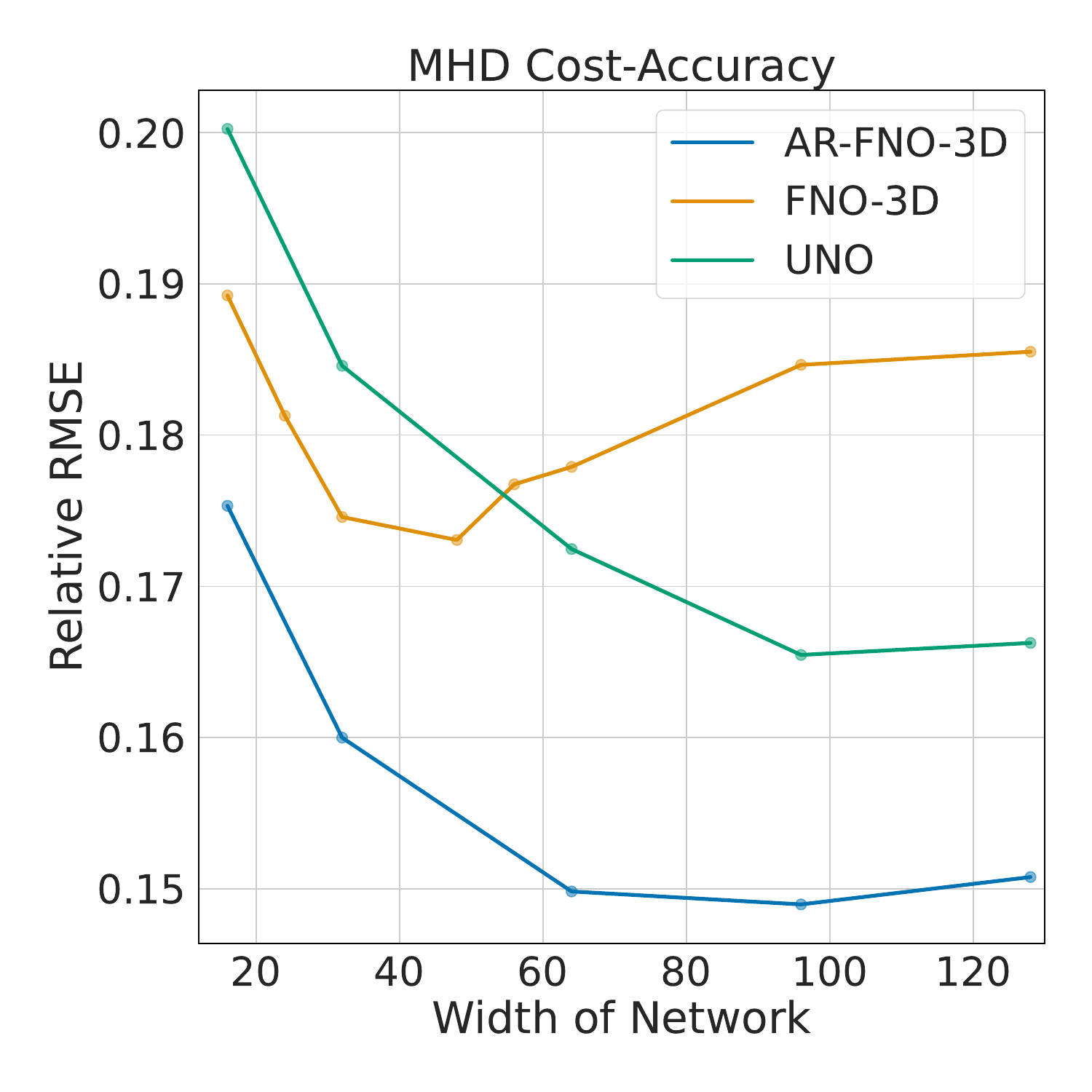}
    \caption{Cost accuracy trade-off for the MHD models. Fieldwise N-RMSE for the best performing models.}
    \label{fig:cost-acc}
\end{figure}

Figure \ref{fig:cost-acc} shows the accuracy of the best-performing models as a function of network width. The auto-regressive FNO-3D's N-RMSE decreases with respect to the width and achieves the lowest relative RMSE. 
The AR-FNO-3D and UNO model errors decrease until a network width of 96 and then overfit at higher widths. The FNO-3D overfits with a width of 32. Due to the relatively early overfitting, we train additional FNO-3D widths of 24, 48, and 56, 
which reveal that the lowest relative RMSE occurs at 48.

The UNO models generally have more parameters at a given network width, allowing the model to better fit the data in fewer epochs. The AR-FNO-3D models have the same number of parameters as the FNO-3D models and train on the intermediate timesteps that FNO-3D  models skip. These timesteps act as additional training data for the AR-FNO-3D architecture.

We compare the inference time of the networks to the Gizmo update time as shown in Table \ref{timing}. We run the models on an NVIDIA A5000 and run the Gizmo simulations on an Intel Xeon Platinum 8160 with 48 cores. 
Table \ref{timing} shows that the neural surrogates are five orders of magnitude faster for the spherical collapse case and one to two orders of magnitude faster for the two turbulence setups. However, these comparisons do not fully account for the differences between these approaches. Gizmo is optimized to run on CPU cores, while Pytorch is optimized for GPUs. Furthermore, our models predict only three physical variables, $(\sigma, u, v)$, on a 3D grid, which acts as proxy for the reduced data availability of observations. We also exclude the training time for the models. In contrast, Gizmo predicts dozens of variables and solves the equations multiple times between each stored snapshot. 
\begin{table*}[!hbt]
    \caption{\textbf{Inference time in seconds to predict five timesteps}.}
    \begin{tabular}{lllll}
      &    &  \textbf{Spherical Collapse }& \textbf{Turbulence} & \textbf{MHD}    \\
         \hline
         Classical Solver &Gizmo       &  758 s& 0.81 s& 1.27 s\\ \hline
         \multirow{4}*{Neural Surrogates}
         & AR-FNO-3D      &  0.626 s& 3.350 s& 3.987 s\\
         &FNO-3D      &  1.130e-02 s& 3.019e-03 s& 2.554e-02 s \\
         &UNO         &  1.786e-02 s& 5.708e-03 s& 5.267e-03 s \\
    \end{tabular}
    \label{timing}
\end{table*}

\section{Ablation Study on Compressibility}

To quantify the effects of the larger dynamic range on model performance, we run an ablation study with the two MHD simulations as shown in Table \ref{RMSE_lowVP}.  We create new training sets that reduce the integrated density dynamic range from $[10^{-3}, 10^{-1}]$ $\frac{g} {cm^{2}}$ to $[8 \times 10^{-3}, 8 \times 10^{-2}]$ $\frac{g} {cm^{2}}$. These simulations adopt a lower virial parameter of 0.12, which corresponds to one-tenth of the turbulent energy of the original MHD setup. 

\begin{table*}
    \caption{Ablation study comparing the single-mapping N-RMSE for each field between the two dynamic ranges. The errors reported are N-RMSE for integrated density, $\delta_\sigma$, N-RMSE for the first velocity component, $\delta_u$, and second velocity component, $\delta_v$, and the average of the errors $\delta_{avg}$. }
    \begin{tabular}{lllllll}
        \hline
        \textbf{Simulation}&\textbf{Method }& \# Parameters& $\delta_{\sigma}$ & $\delta_u$ & $\delta_v$ & $\delta_{avg}$ \\ \hline
        &Identity& N/A&  0.435&  0.268&  0.310&  0.338\\
        \textbf{Higher Dynamic Range}  &AR-FNO-3D& 1.55e+08&  0.306&  0.174&  0.209&  0.229 \\
        $[10^{-3}, 10^{-1}]$ $\frac{g} {cm^{2}}$&FNO-3D& 6.54e+07&  0.234&  0.132&  0.156&  0.174\\
        &UNO& 1.40e+07&  0.231&  0.132&  0.157&  0.173\\ \hline
        &Identity& N/A&  0.088&  0.123&  0.107&  0.106 \\
        \textbf{Lower Dynamic Range} &AR-FNO-3D& 1.17e+08&  0.038&  0.069&  0.058&  0.055\\
        $[8 \cdot 10^{-3}, 8 \cdot 10^{-2}]$ $\frac{g} {cm^{2}}$ &FNO-3D& 6.54e+07&  0.037&  0.057&  0.047&  0.047\\
        &UNO& 1.40e+07&  0.032&  0.050&  0.041&  0.041 \\ \hline
    \end{tabular}
    \label{RMSE_lowVP}
\end{table*}
We note that reducing the effective Mach number increases the characteristic dynamical time ($t_{\rm dym} = L / v_{\rm rms}$). Consequently, the change in the gas state between snapshots is smaller for the same output interval.
In this case, the target operator is closer to the identity mapping. To address this, we calculate the N-RMSE between the 
predicted output and the input data ( the identity operator).
The lower compressibility reduces the error by approximately a factor of four for all network architectures. 

\section{Single Observational Experiment}

Our experimental setup trains from sequential time series, which are unavailable for astronomical observations. 
To explore the impact of reduced data, we conduct a test with a single input observation, i.e., we use the same time snapshot for all five steps in the input series. This allows us to test our trained model in this low-information scenario.
We forecast the initial condition repeated multiple times to make a ``temporal" function with the best-performing models as done in a prior study
\cite{wang2024closuremodelslearningchaoticsystems}. We run this experiment with the models trained on the MHD dataset, since these physics are closest to actual observations. 

\begin{figure*}[!hbt]
\centering
\begin{center}

  \includegraphics[width=0.85\linewidth]{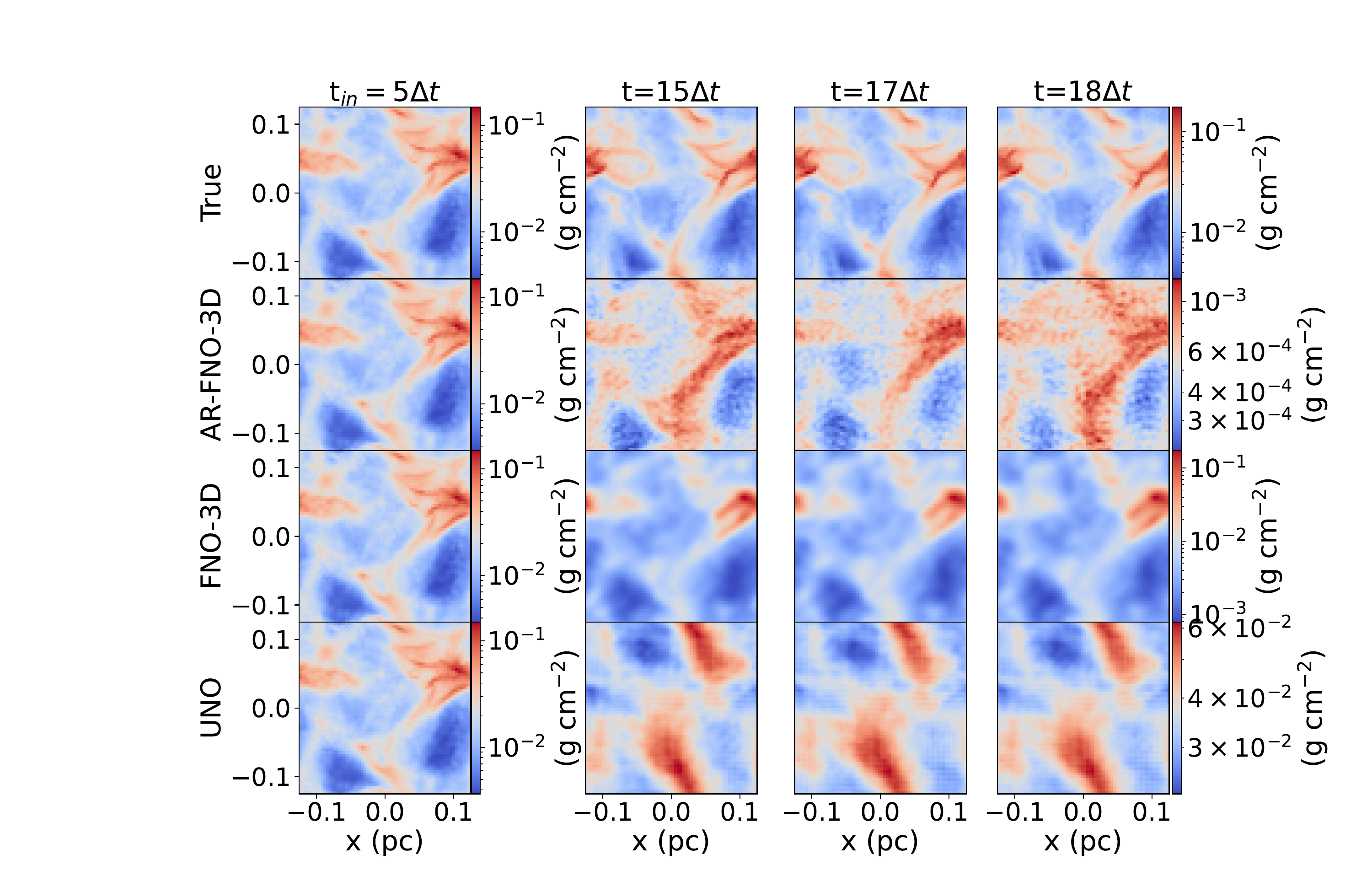}
  \caption{Temporal evolution from left to right of the single-observation input test. The first column is the initial condition input (input five times) into the model, while each subsequent column depicts the first ($t=15\Delta t$), third ($t=17\Delta t$), and final ($t= 18 \Delta t$) output times to show the evolution in the predictions. The timestep is $\Delta t =$8.29 kiloyears. The rows from top to bottom are the Gizmo solution, the AR-FNO-3D prediction, the FNO-3D prediction, and the UNO prediction, respectively. Each model has a different scale for the colorbar to allow comparisons between the model performance.}\label{SingleObs}
  \end{center}
\end{figure*}

Figure \ref{SingleObs} depicts the forecasted initial conditions of the model. Since we train the models on a timeseries, the model outputs should evolve temporally even though the inputs do not evolve. Both the FNO-3D and the UNO exhibit only small differences and miss much of the small-scale structure. We find that the intensity of the predictions decreases significantly in some of the models, e.g., the AR-FNO-3D model. To better enable comparisons between the models, we adjust the Figure \ref{SingleObs} colorbars so that the predicted structure is visible in all panels.

The UNO model predicts two large masses, which do not resemble the ground truth and exhibit a lower peak mass density and smaller dynamic range. The FNO-3D model performs the best, and its predictions span the expected dynamic range. The AR-FNO-3D model loosely emulates the structure of the ground truth, however, the density spans a very small range of $[10^{-4}, 10^{3}]$. 



\newpage
\section{Additional Figures for the Spherical Collapse Setup.}\label{Spherical_App}

\textbf{The figures below show the velocity predictions for the spherical collapse models.}\label{App_Grav_FNO3d}
\begin{figure}[!hbt]
\centering
\begin{center}

  \includegraphics[width=0.85\linewidth]{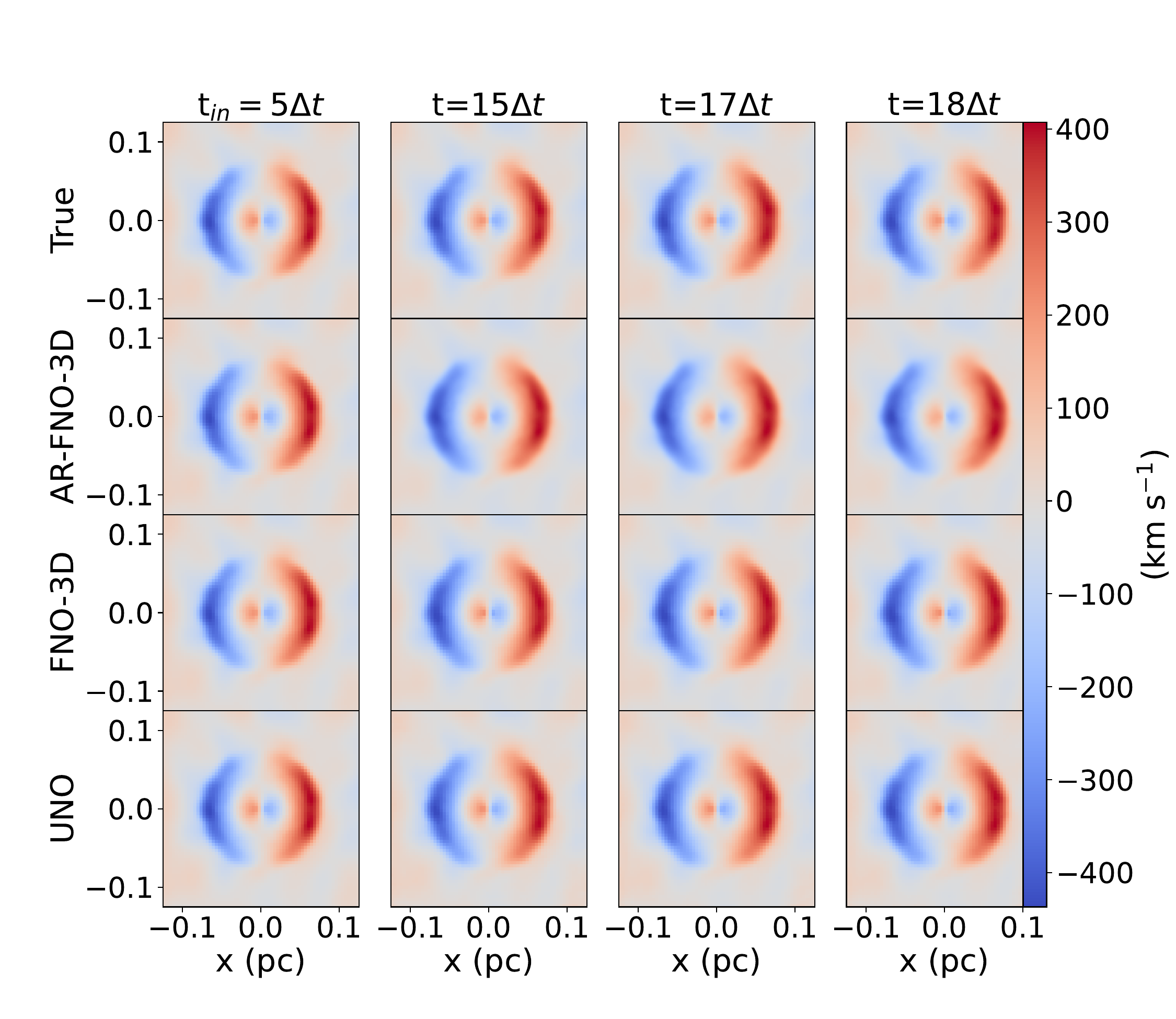}
  \caption{Temporal evolution from left to right of the first projected velocity component for a collapsing sphere. The first column is the final timestep input into the model, while each subsequent column depicts the first ($t=15\Delta t$), third ($t=17\Delta t$), and final ($t= 18 \Delta t$) output times to show the evolution in the predictions. In this snapshot, the initial sphere mass is $0.63 M_{\odot}$ and the timestep, $\Delta t =$0.18 kilo-years. The rows from top to bottom are the Gizmo solution, the AR-FNO-3D prediction, the FNO-3D prediction, and the UNO prediction, respectively.}\label{Grav_vx}
  \end{center}
\end{figure}
\newpage
\begin{figure}[!hbt]
\centering
\begin{center}

  \includegraphics[width=0.85\linewidth]{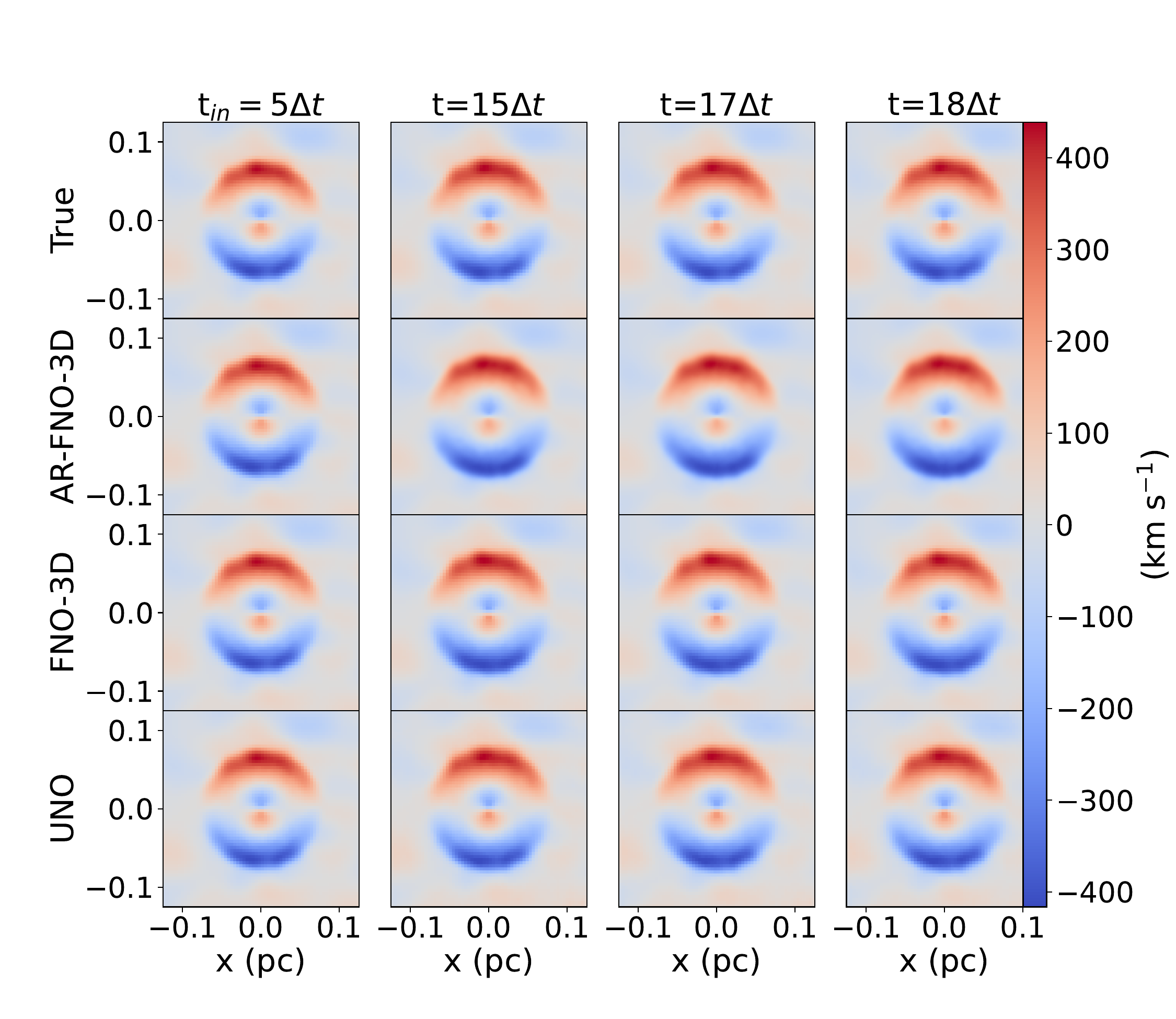}
  \caption{Temporal evolution from left to right of the second projected velocity component for a collapsing sphere. The first column is the final timestep input into the model, while each sequential columns depict the first ($t=15\Delta t$), the third ($t=17\Delta t$), and the final ($t= 18 \Delta t$) output times to show the evolution in the predictions. In this snapshot, the initial sphere mass is set to $0.63 M_{\odot}$ and the timestep, $\Delta t =$0.18 kilo-years. The rows from top to bottom are the Gizmo solution, the AR-FNO-3D prediction, the FNO-3D predictions, and the UNO predictions, respectively.}\label{Grav_vy}
  \end{center}
\end{figure}

\newpage
\section{Plots For Turbulent Flow} \label{Turb_App}
\textbf{Velocity predictions for Turbulent Flow.}
\begin{figure}[!hbt]
\centering
\begin{center}

  \includegraphics[width=0.85\linewidth]{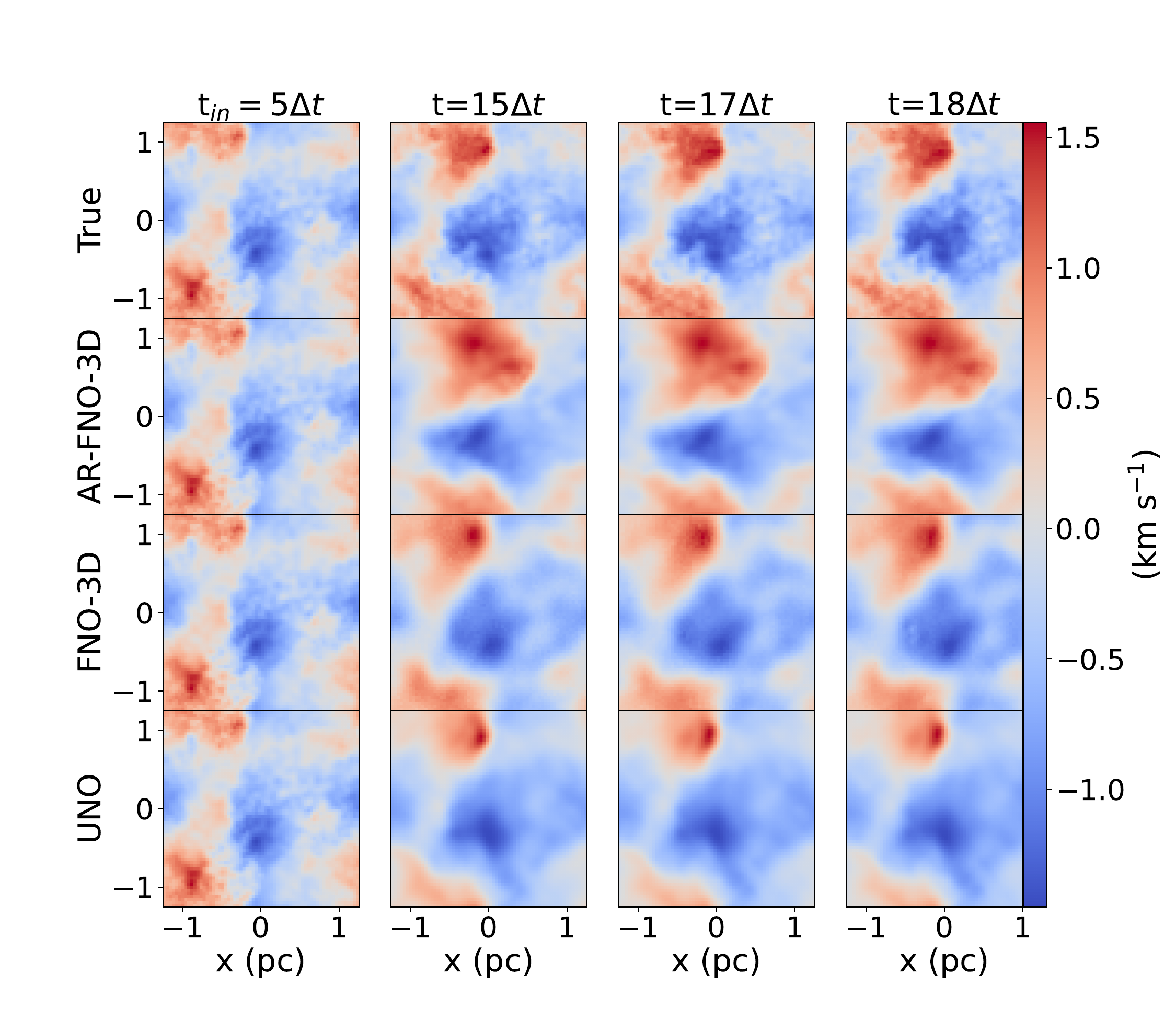}
  \caption{Temporal evolution from left to right of the first projected velocity component after the fluid reaches a quasi-steady turbulence state. The first column is the final timestep input into the model, while each sequential columns depict the first ($t=15\Delta t$), the third ($t=17\Delta t$), and the final ($t= 18 \Delta t$) output times to show the evolution in the predictions. The rows from top to bottom are the Gizmo solution, the AR-FNO-3D prediction, the FNO-3D predictions, and the UNO predictions, respectively.}\label{turb_vx}
  \end{center}
\end{figure}
\newpage
\begin{figure}[!hbt]
\centering
\begin{center}

  \includegraphics[width=0.85\linewidth]{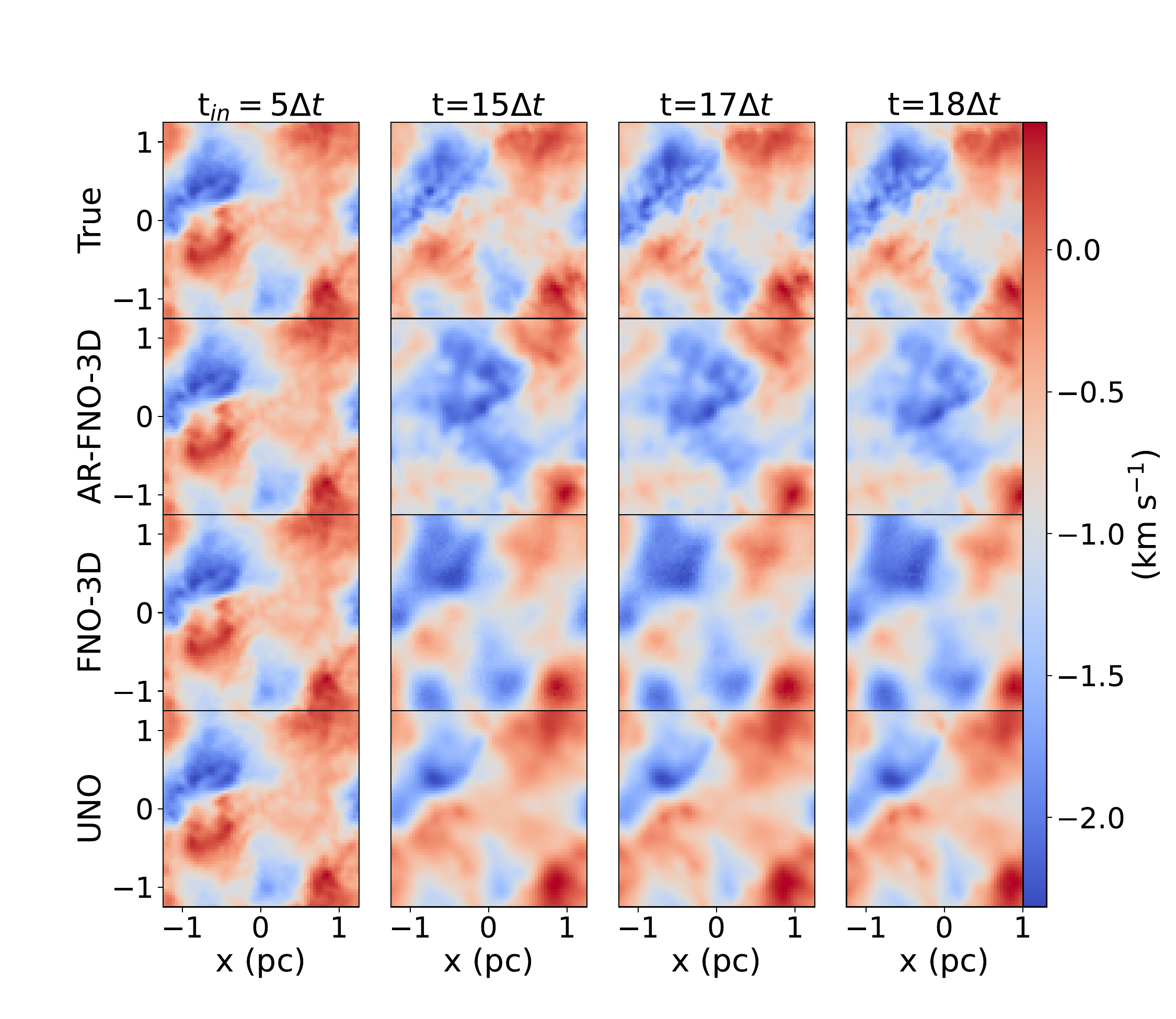}
  \caption{Temporal evolution from left to right of the second projected velocity component after the fluid reaches a quasi-steady turbulence state. The first column is the final timestep input into the model, while each sequential columns depict the first ($t=15\Delta t$), the third ($t=17\Delta t$), and the final ($t= 18 \Delta t$) output times to show the evolution in the predictions. The rows from top to bottom are the Gizmo solution, the AR-FNO-3D prediction, the FNO-3D predictions, and the UNO predictions, respectively.}\label{turb_vy}
  \end{center}
\end{figure}

\section{Plots for MHD Turbulunce} \label{MHD_App}
\textbf{Velocity predictions for Turbulent magnetohydrodynamics.}
\begin{figure*}[!hbt]
\centering
\begin{center}
  \includegraphics[width=0.85\linewidth]{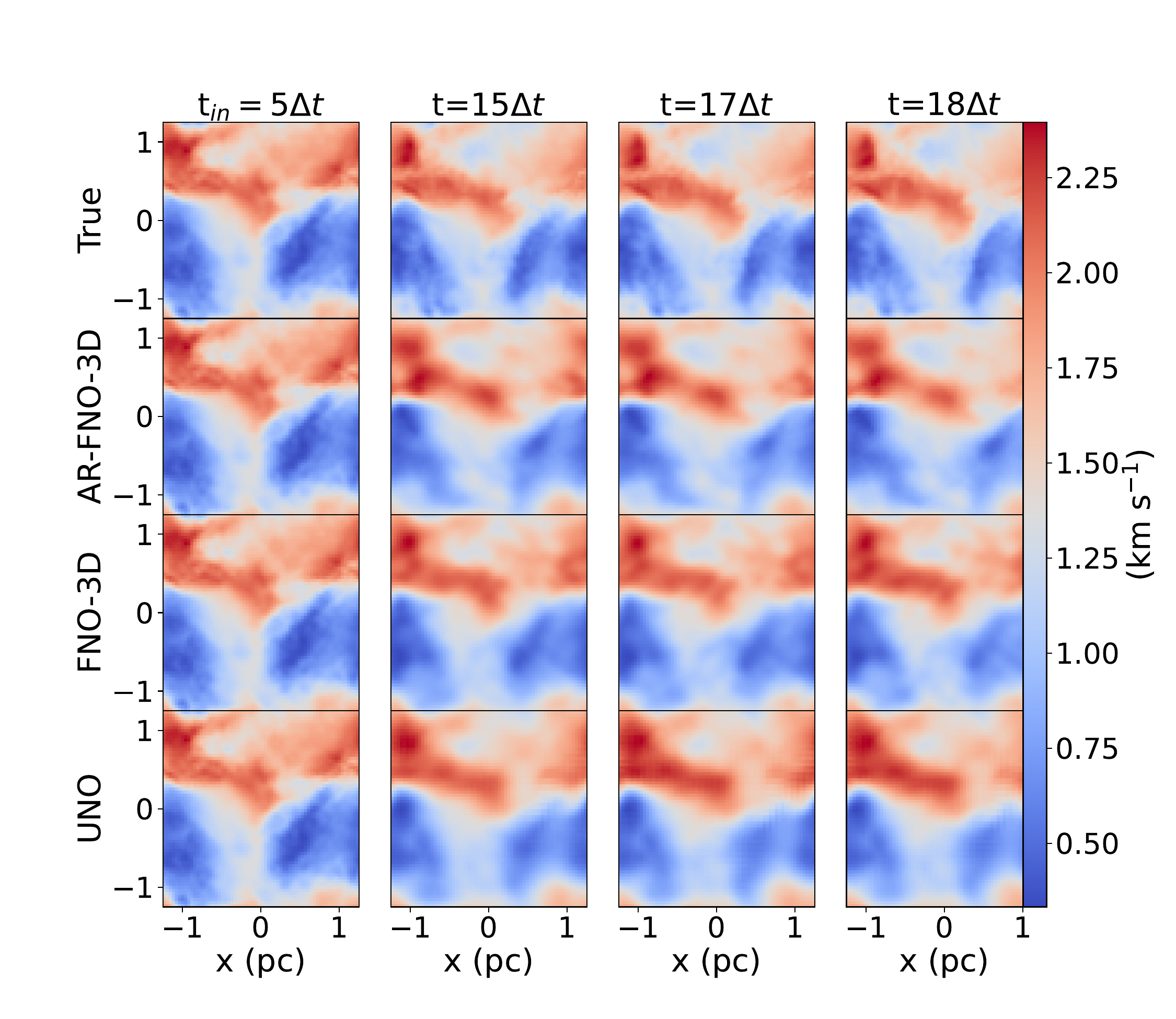}
  \caption{Temporal evolution from left to right of the first projected velocity component after the magnetized fluid reaches a quasi-steady turbulence state. The first column is the final timestep input into the model, while each sequential columns depict the first ($t=15\Delta t$), the third ($t=17\Delta t$), and the final ($t= 18 \Delta t$) output times to show the evolution in the predictions. The rows from top to bottom are the Gizmo solution, the AR-FNO-3D prediction, the FNO-3D predictions, and the UNO predictions, respectively.}\label{mag_vx}
  \end{center}
\end{figure*}
\begin{figure}[!hbt]
\centering
\begin{center}

  \includegraphics[width=0.85\linewidth]{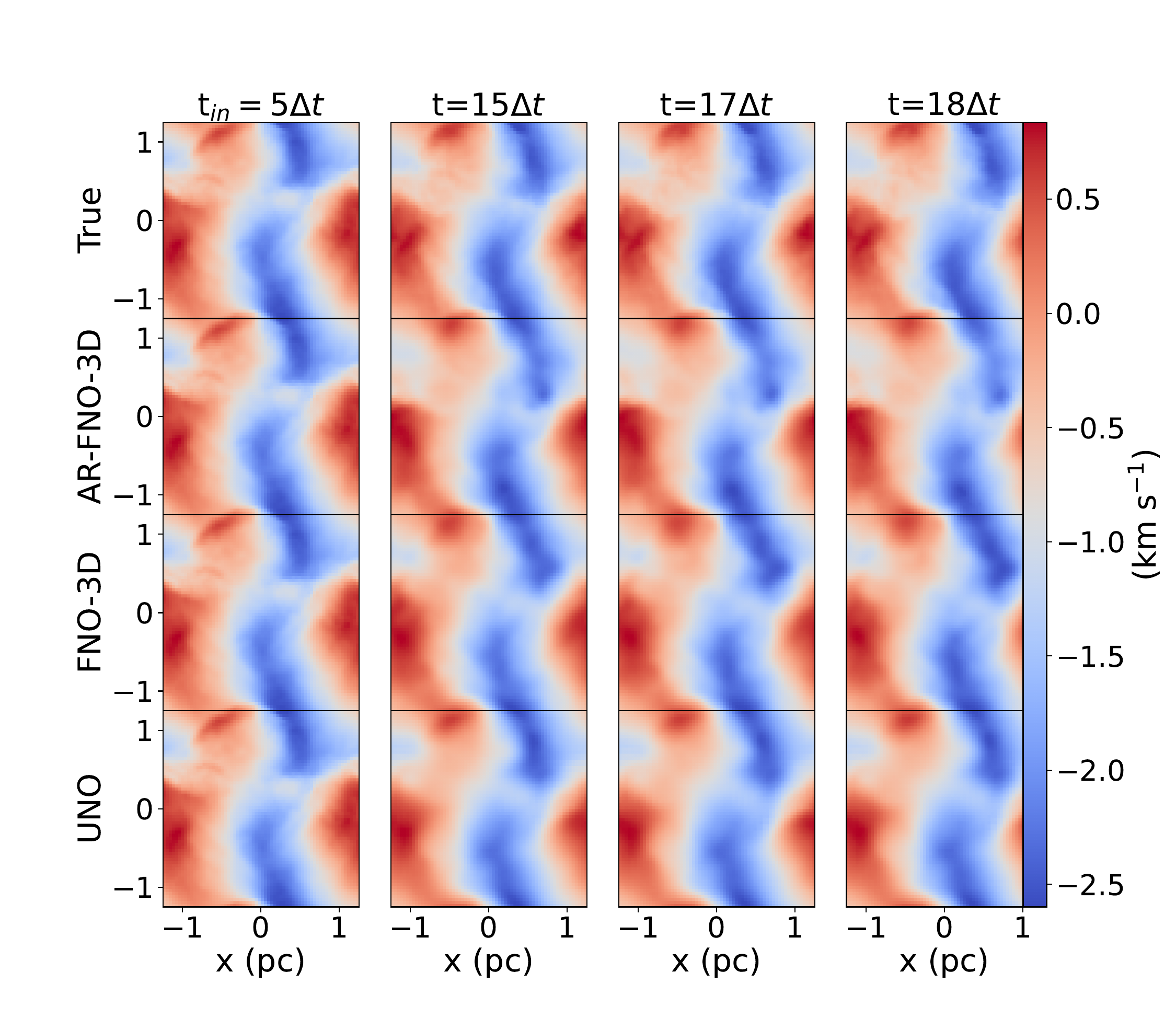}
  \caption{Temporal evolution from left to right of the second projected velocity component after the magnetized fluid reaches a quasi-steady turbulence state. The first column is the final timestep input into the model, while each sequential columns depict the first ($t=15\Delta t$), the third ($t=17\Delta t$), and the final ($t= 18 \Delta t$) output times to show the evolution in the predictions. The rows from top to bottom are the Gizmo solution, the AR-FNO-3D prediction, the FNO-3D predictions, and the UNO predictions, respectively.}\label{mag_vy}
  \end{center}
\end{figure}
\bibliography{main}
\end{document}